\documentstyle[a4,12pt,psfig,twoside]{article}
\begin{document}
\newcommand{\beq}{\begin{equation}}
\newcommand{\eeq}{\end{equation}}
\newcommand{\ie}{{\em i.e. }}
\newcommand{\al}{{\em et al }}
\newcommand{\eg}{{\em e.g. }}
\title{Cluster Analysis of Gene Expression Data \thanks{
Paper presented in celebration of Michael Fisher's seventieth birthday;
knowing Michael, learning from him and arguing with him have been among
the greatest privileges of my
professional life.}
}

\author{Eytan Domany \\
Department of Physics of Complex Systems, Weizmann Institute of
Science, \\ Rehovot 76100, Israel\\
%E-mail: eytan.domany@weizmann.ac.il
}

%%%%%%%%%%%%%%%%%%%%%%%%%%%%%%%%%%%%%%%%%%%%%%%%%%%%%%%%%%%%%%
% You may repeat \author \address as often as necessary      %
%%%%%%%%%%%%%%%%%%%%%%%%%%%%%%%%%%%%%%%%%%%%%%%%%%%%%%%%%%%%%%
\date{\today}
\maketitle

\begin{abstract}
The expression levels of many thousands of genes can be measured 
simultaneously
by DNA microarrays (chips). This
novel experimental tool has revolutionized
research in molecular biology and generated considerable excitement.
A typical experiment uses a few tens of such chips,
each dedicated to a single  sample - such as tissue extracted from
a particular tumor. The results of such an experiment contain several
hundred thousand numbers, that come in the
form of a table, of several thousand rows (one for each gene) and
50 - 100 columns (one for each sample). We developed a clustering 
methodology
to mine such data. In this review
I provide a very basic introduction to the subject, aimed at a physics 
audience
with no prior knowledge of either gene expression
or clustering methods. I explain what genes are, what is gene expression
and how it is measured by DNA chips. Next I explain what is meant by
"clustering" and how we analyze the massive amounts of data from
such experiments, and present results obtained from analysis of data
obtained from colon cancer, brain tumors and breast cancer.
\end{abstract}

\section{Reflection and Outline }
The subject of this paper does not seem to have much to do with Statistical
Mechanics, the subject I learned from Michael Fisher. Indeed the aim of 
the research I am describing is to gain understanding of {\it Biology},
and the methodology used is in the realm of {\it Applied Mathematics} and
{\it Pattern Recognition}. Closer inspection reveals, however, that the
ideas that underlie the approach rely strongly on the very subjects to which
I have been introduced by Michael: Monte Carlo simulations\cite{MC} and phase
transitions in Potts ferromagnets\cite{Potts}. The problem area and technology
I describe below are among of the most fascinating and exciting topics I 
encountered. I hope that Michael, who always had a keen interest in biology,
will find the applicability of Statistical Physics to this type
of research gratifying.  

This paper has three parts, aimed at explaining the
meaning of it's title. The first part is a telegraphic introduction to
the relevant biology, starting from genes and transcription and ending with 
an explanation of what  DNA chips are and the kind of data that they 
produce.
The second part is an equally
concise introduction to cluster analysis, leading to a recently
introduced method, Coupled Two-Way Clustering (CTWC), that was designed
for the analysis and mining of data obtained by DNA chips. The third
section puts the two introductory parts together and demonstrates
how CTWC is used to obtain insights from the anaysis of gene expression
data in several clinically relevant contexts, such as colon cancer,
glioblastoma and breast cancer.

\section{Introduction to the relevant biology}

\subsection{Genes and Gene Expression}
Since my aim is to introduce only
those concepts that are
absolutely essential for understanding the data that will be
presented and analyzed, I present here only a severely oversimplified
description of
a large number of very complex processes.
The interested reader is referred to two excellent
textbooks~\cite{Alberts,Gould}.
%%%%%%%%ad can
\begin{figure}[t]
%\figurebox{20pc}{15pc}{} % to have a box alone
%\epsfxsize=10pc % will enlarge or reduce the postscript figures based 
%on the xsize
\centerline{
\psfig{figure=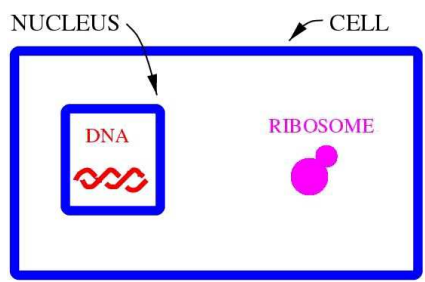,height=4cm}
}
\caption{Caricature of a
eucaryotic cell: its nucleus contains DNA, whereas the ribosomes
are in the cytoplasm.
\label{fig:cell}}
\end{figure}
%%%%%%%%%%%%%%%%%%%%
Cells and organisms are divided into two classes; procaryotic (such as 
bacteria)
and eucaryotic. The latter have a nucleus; see the schematic drawing 
of
Fig \ref{fig:cell}. The cell is
enclosed by
it's membrane;
embedded in the
cell's  cytoplasm
is it's nucleus,
surrounded and protected by its own membrane.  The nucleus contains 
DNA, a one dimensional molecule, made of two complementary strands,
coiled around each other as a double helix.
Each strand consists of a
backbone to which a linear sequence of bases is attached. There are four
kinds of bases, denoted by C,G,A,T. The two strands contain complementary
base sequences and are held together by hydrogen
bonds that connect matching pairs of bases; G-C (three hydrogen bonds) and 
A-T (two).

A {\it {\bf gene} is a segment of DNA, which contains the formula
for the chemical
composition of one particular protein}. Proteins are the working molecules of
life; nearly every biological function is carried out by a protein.
Topologically, a protein
is also a chain; each link is an amino acid, with neighbors along the chain
connected by covalent bonds. All proteins are made of 20 different amino
acids - hence the chemical formula of a protein of length $N$ is an $N$-letter
word, whose letters are taken from a 20-letter alphabet.
A gene is nothing but an alphabetic cookbook recipe, listing
the order in which the amino acids are to be strung when the corresponding
protein is synthesized.
Genetic information is encoded in the
linear sequence in which the bases on the two strands
are ordered along the DNA molecule. The {\it genetic code} is a universal
translation table, with specific triplets of consecutive bases coding for every
amino acid.

The {\it {\bf genome} is the collection of all the
chemical formulae for the proteins that an organism needs and produces}.
The genome of a simple
organism such as yeast contains about 7000 genes;
the human genome has between
30,000 and 40,000. An overwhelming majority
(98\%) of human DNA contains non-coding
regions (introns), i.e. strands that do not code for any particular protein.

Here is an amazing fact; every cell of a multicellular organism contains its
entire genome! That is, every cell has the entire set of recipes the organism
may ever need; the nucleus of each of
the reader's cells contains every piece of information
needed to make a copy (clone) of him/her! Even though each cell contains the same
set of genes, there is {\it differentiation:}
cells of a complex organism, taken from different
organs, have entirely different functions and the proteins that
perform these functions are very different. Cells in our retina
need photosensitive molecules, whereas our livers do not make
much use of these. A {\bf gene is
expressed} {\it in a cell when the protein it codes for is actually
synthesized}. In an average human cell about 10,000 genes are expressed.

The large majority of abundantly expressed genes are associated with common
functions, such as metabolism, and hence are expressed in all cells. However,
there will be differences between the expression profiles of different cells,
and even in a single cell, expression will vary with time, in a manner dictated
by external and internal signals that reflect the state of the organism and the
cell itself.
%Hence even though all the formulae are in every cell, in
%a particular cell only a subset of the genes are expressed.

Synthesis of proteins takes place at the {\it \bf ribosomes}. These are enormous
machines (made also of proteins) that read the chemical formulae written
on the DNA and synthetise the proten according to the instructions.
The ribosomes are in the cytoplasm,
whereas the DNA is in the protected environment of the nucleus. This poses
an immediate logistic problem - how does the information get transferred
from the nucleus to the ribosome?

\subsection{Transcription and Translation}
The obvious solution of information transfer would be to rip out the
piece of DNA that contains the gene that is to be expressed, and transport
it to the cytoplasm. The engineering analogue of this strategy is the
following. Imagine an architect, who has a single copy of a design for a
building, stored on the hard disk of his PC. Now he has to transfer the
blueprint to the construction site, in a different city. He probably will
not opt for tearing out his hard disk and mailing it to the site, risking
it being irreversibly lost or corrupted. Rather,
he will prepare several diskettes, that contain copies of his design,
and mail these in separate envelopes.

This is precisely the strategy adopted by cells.
\begin{figure}
  \centerline{\psfig{figure=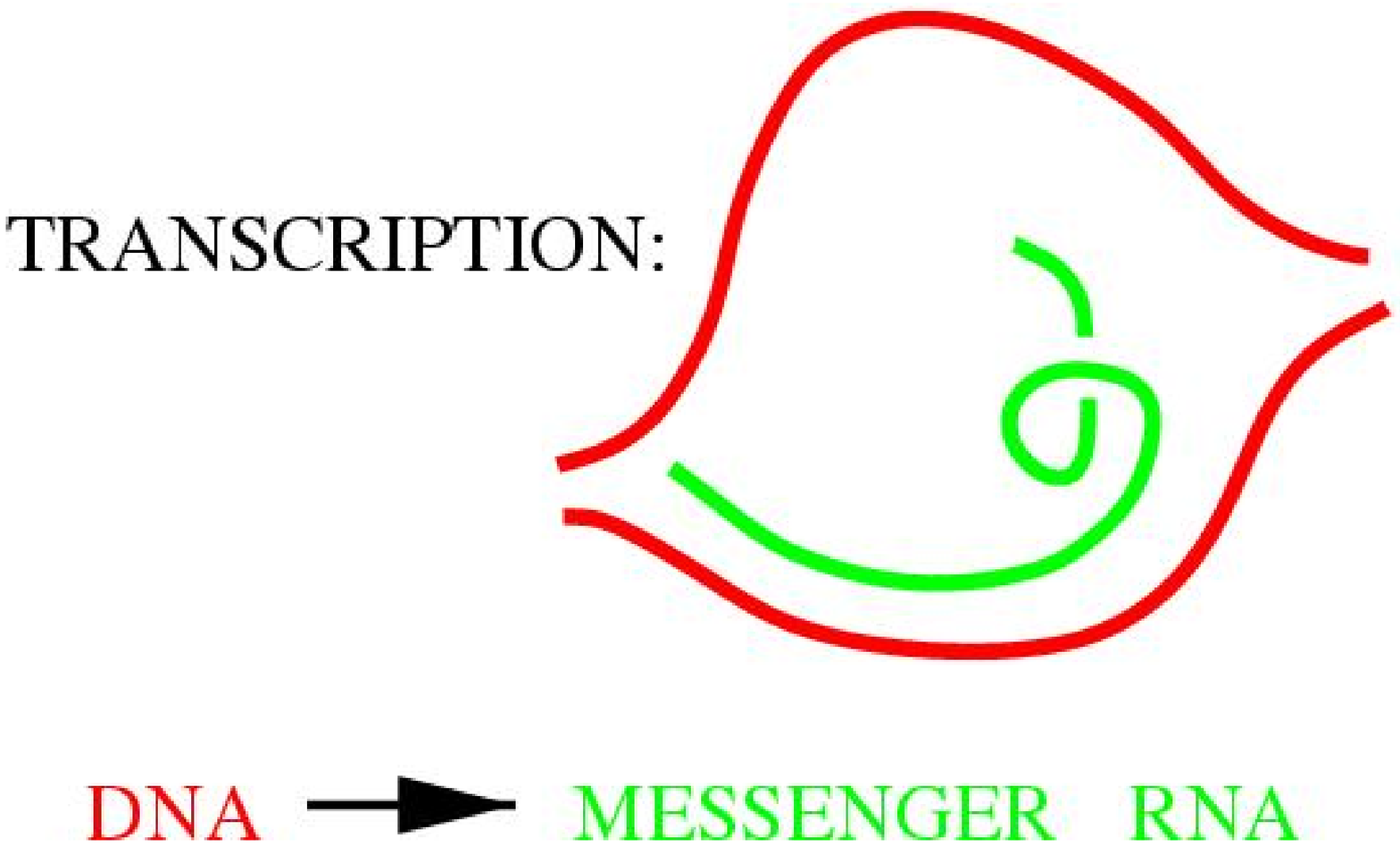,height=4cm}
    \hspace{1.0cm}
    \psfig{figure=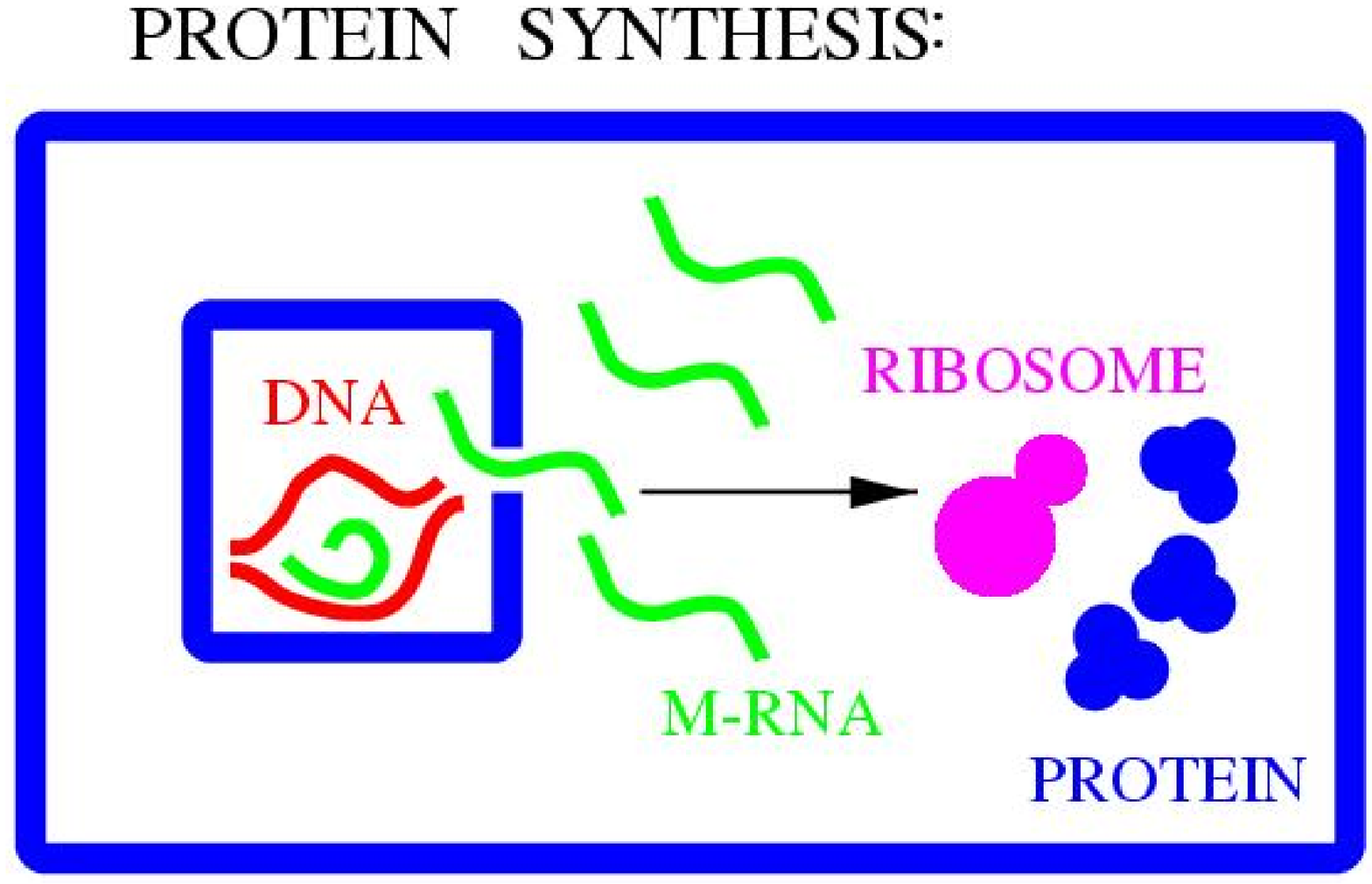,height=4cm}
    }
  \caption{Transcription involves synthesis of mRNA, a copy of the
  gene encoded on the DNA (left). The mRNA molecules leave the nucleus
  and serve as the template for protein synthesis by the ribosomes (right).}
  \label{fig:trans}
\end{figure}

When a gene receives
a command to be expressed, the corresponding double helix of DNA opens, and
a precise copy of the information, as written on one of the strands,
%%%%%%%%%%%%%%%%%%%%%%%%%%%%%%%%%%
is prepared (see Fig \ref{fig:trans}). This "diskette" is a linear molecule
%%%%%%%%%%%%%%%%%%%%%%%%%%%%%%%%%%
called {\it \bf messenger
RNA, mRNA} and the process of its production is called {\it transcription}.
The  subsequent reading of the mRNA, deciphering the message (written using base
triplets) into amino acids
and synthesis of the corresponding protein at the ribosomes
\footnote{Actually the mRNA is "read" by one end of another molecule,
transfer RNA; the amino acid that corresponds to the triplet of bases
that has just been read is attached to the other end of the tRNA. This process,
and the formation of the peptide bond between subsequent amino acids, takes
place on the ribosome, which moves along the mRNA as it is read.}
is called {\it translation}. In fact, when many molecules of
a certain protein are needed, the cell produces many corresponding mRNAs,
which are transferred through the nucleus' membrane to the cytoplasm, and are
"read" by several ribosomes. Thus the single master copy of the
instructions,
%%%%%%%%%%%%%%%%%%%%%%%%%%%%%%%%%%%%%%%%%%%%%%%%%%%%%%
contained in the DNA, generates many copies of the protein (see Fig
\ref{fig:trans}).
%%%%%%%%%%%%%%%%%%%%%%%%%%%%%%%%%%%%%%%%%%%%%%%%%%%%%%
This transcription  strategy is prudent and safe,
preserving the precious master copy;
at the same time it also serves as
a remarkable amplifier of the genetic information.

A cell may need a large number of some proteins and a small number of others.
That is, every gene may be expressed at a different level.
The manner in which the instructions to start and stop transcription are
given for a certain gene
%and to when the desired concentration of the coded protein
%is reached
is governed by {\it regulatory networks}, which constitute one of the most
intricate and fascinating subjects of current research.
%which lies beyond the scope of this
%crash course. and will be touched upon only briefly.
Transcription is regulated
by special proteins, called {\it transcription factors},
which bind to specific locations on the DNA, upstream from the coding
region. Their presence at the right site initiates or suppresses transcription.

This leads us to the {\it basic paradigm of gene expression analysis}:
\begin{quote}
The "biological state" of a cell is reflected by its {\it expression profile}:
the expression levels of all the genes of the genome. These, in turn,
are reflected by the concentrations of the corresponding mRNA molecules.
\end{quote}

%%%ad can

This paradigm is by no means trivial or perfectly true. One may argue that
the state of a cell at a given moment
is defined by its chemical composition, i.e. the concentration of all
the constituent proteins. There is no assurance that these concentrations
are directly proportional to the concentrations of the related mRNA
molecules. The rates of degradation of the different
mRNA, the efficiency of their transcription
to proteins, the rate of degradation of the proteins - all these may
vary. Nevertheless,
this is our working assumption; specifically, we assume that for human cells
the expression levels of all 40,000 genes completely specify the state
of the particular tissue from which the cells were taken.
The question we turn to answer is - how does one measure,
for a given cell or tissue,
the expression levels of thousands of genes?

\subsection{DNA chips}

A {\it {\bf DNA chip } is the instrument that measures simultaneously
the concentration of thousands of different mRNA
molecules}. It is also referred to as a DNA microarray
(see~\cite{Schulze01} for a recent review of the
technology, and the special supplement of {\it Nature Genetics} 
{\bf 21}, Jan. 1999).
DNA microarrays, produced by
Affymetrix\cite{Affy}, can
measure simultaneously the expression levels of up to 20,000 genes;
the less expensive
spotted arrays\cite{spot} do the same for several thousand. Schematically, 
the Affymetrix
arrays are produced as follows.
Divide
a chip (a glass plate of about 1 cm across) into "pixels" - each
dedicated to one gene {\it g}.
Billions of 25 base pair long pieces (oligonucleotides)
of {\it single strand DNA}, copied from a particular segment of gene {\it g}
are photolitigraphically synthesised on the dedicated pixel
(these are referred to as "probes")\footnote{
Actually next to a pixel of 25-mers that are perfect copies of a bit of a gene, 
one
places copies of {\it mismatched} 25-mers - in these a central base has been 
changed.
One then measure the difference between hybridization to perfect match (PM) and 
mismatch (MM). Each gene is represented on a chip by 20 such pairs of 25-mers.}
.
The mRNA molecules are extracted from the
cells taken from the tissue of
interest (such as tumor tissue obtained by surgery). They are {\it Reverse 
Transcribed}
from RNA to DNA and their concentration is
enhanced. Next, the resulting DNA is transcribed back into
fluorescently marked single strand RNA.
The solution of marked and enhanced mRNA molecules ("targets") that are copies 
of the
mRNA molecules that were  originally extracted from the tissue,
is placed on the chip
and the labeled RNA are
diffusing over the dense
forest of single strand DNA probes. When such an mRNA
encounters a part of the gene of which it is a perfect copy, it attaches to
it - {\it hybridizes} - with a high affinity
(considerably higher than with a bit of DNA of
which it is not a perfect copy). When the mRNA solution is washed off, only
those molecules that found their perfect match remain stuck to the chip.
Now the chip is illuminated with a laser, and these stuck "targets" fluoresce;
by measuring the light intensity emanating from each pixel, one obtains a
measure of the number of targets that stuck, which, in turn, is proportional
to the concentration of these mRNA in the investigated tissue. In this manner
one obtains, from a chip on which $N_g$ genes were placed, $N_g$ numbers
that represent the expression levels of these genes in that tissue.
A typical experiment provides the expression
profiles of several tens of samples (say $N_s \approx 100$), over several
thousand ($N_g$) genes.
These results are summarized in an $N_g \times N_s$ {\it expression table};
each row corresponds to one particular gene and each column to a sample.
Entry $E_{gs}$
of such an expression table stands for the expression level of gene $g$ in
sample $s$. For example, the experiment on colon cancer, first reported
by Alon et al~\cite{Alon99}, contains $N_g=2000$ genes whose expression
levels passed some
threshold, over $N_s=62$ samples, 40 of which were taken from tumor and 22
from normal colon tissue.

%%%ad can

Such an expression table contains up to several hundred thousand numbers;
the main issue addresed in this paper concerns the manner in which
such vast amounts of data are "mined", to extract from it
biologically relevant
meaning.
%%%%%%%%%%%%%%%ad can
Several obvious aims of the data analysis are the following:
\begin{enumerate}
\item
Identify genes whose expression levels reflect biological processes
of interest (such as development of cancer).
\item
Group the tumors ito classes that can be differentiated on the basis of
their expression profiles, possibly in a way that can be interpreted
in terms of clinical classification. If one can partition tumors,
on the basis of their expression levels, into relevant classes (such as
e.g. positive vs negative responders to a particular treatment), the
classification obtained from expression analysis can be used as a
diagnostic and thereupeutic tool\footnote{For example one hopes to
use the expression profile of a tumor to select the most effective therapy.}.
\item
Finally, the analysis can provide clues and guesses for the function of
genes (proteins) of yet unknown role\footnote{The statement "the human genome
has been solved" means that the sequences of 40,000 genes are known, from which
the chemical formulae of 40,000 proteins can be obtained.
Their biological function, however, remains largely unknown.}.
\end{enumerate}
This concludes the brief and very oversimplified review of the biology
background
that is essential to understand the aims of this research.
In what follows I present a method designed for mining such expression data.

\section{Cluster Analysis}
\subsection{Supervised versus unsupervised analysis}
Say we have two groups of samples, that have been labeled on the basis
of some external (i.e. not contained in the expression table) information,
such as clinical identification of tumor and normal samples; and
our aim is to identify genes whose expression levels are significantly
different for these two groups.
Supervised analysis is the most suitable method for this kind of task.
The simplest way is to treat the genes one at a time; for gene $g$ we
have $N_s$ expression levels $A_{gs}$, and we
propose as a null hypothesis that the these numbers were
picked at random, from the same distribution, for all samples $s$. There are
well established methods to test the validity of such a hypothesis and to
calculate for each gene a statistic whose value indicates whether the
null hypothesis should be accepted or rejected, as well as the probability
$P_g$ for error (i.e. for rejecting the null hypothesis on the basis of the
data, even though it is correct).
An alternative  supervised analysis uses a subset of the tissues of known
clinical label to train a neural network to separate them into the known classes
on the basis of their expression profiles. The generalization
ability of the network is then estimated by classifying
a test set of samples (whose correct labels are also known),
that was not used in the training process.

The main disadvantage of supervised methods is their being {\it limited to
hypothesis testing}. If one has some prior knowledge which can lead to
a hypothesis, supervised methods will help to accept or reject it. {\it They will
never reveal the unexpected and never lead to new hypotheses}, or
to new partitions
of the data. For example, if the tumors break into two unanticipated classes
on the basis of their expression profiles, a supervised method will not
be able to discover this. Another shortcoming is the (often very common)
possibility of misclassification of some samples. A supervised method will
not discover, in general, samples that were mistakenly labeled and used in, say,
the training set.

The alternative is to use {\it unsupervised methods of analysis}. These aim at
exploratory analysis of the data, introducing as little external knowledge
or bias as possible, and "let the data speak". That is, we explore the
structure of the data on the basis of correlations and similarities that are
present in it. In the context of gene expression, such analysis has two
obvious goals:
\begin{enumerate}
\item
Find groups of genes that have correlated expression profiles. The members
of such a group may take part in the same biological process.
\item
Divide the tissues into groups with similar gene expression profiles. Tissues
that belong to one group are expected to be in the same biological
(e.g. clinical) state.
\end{enumerate}
The method presented here to accomplish these aims is called {\it clustering}.

\subsection{Clustering - statement of the problem.}
The aims of cluster analysis~\cite{Jain,Duda} can be stated as
follows: given
$N$ data points, $\mbox{\boldmath X}_i,~~i=1,...,N$ embedded in $D$-dimensional space
(i.e. each point is represented by $D$ components or coordinates),
{\it identify the underlying structure of the data}. That is, peartition
the $N$ points into $M$ {\it \bf clusters}, such that points that belong
to the same cluster are "more similar" to each other than two points that
belong to different clusters. In other words, one aims to determine whether
the $N$ points form a single "cloud", or two, or more; in respectable
unsupervised methods the number of clusters, $M$, is also determined by
the algorithm.

The clustering problem, as stated above, is clearly ill posed. No definition
was given for what is "more similar"; furthermore,
as we will see, the manner in which data points are
assigned to clusters depends on the {\it resolution} at which the data are
viewed. The last concern is addressed by generating a {\it dendrogram}
or tree of clusters, whose number and composition varies with
the resolution that is used. To clarify these points I present a simple example
for a process of "learning without a teacher", of which clustering constitutes
a particular case.

\begin{figure}
  \centerline{\psfig{figure=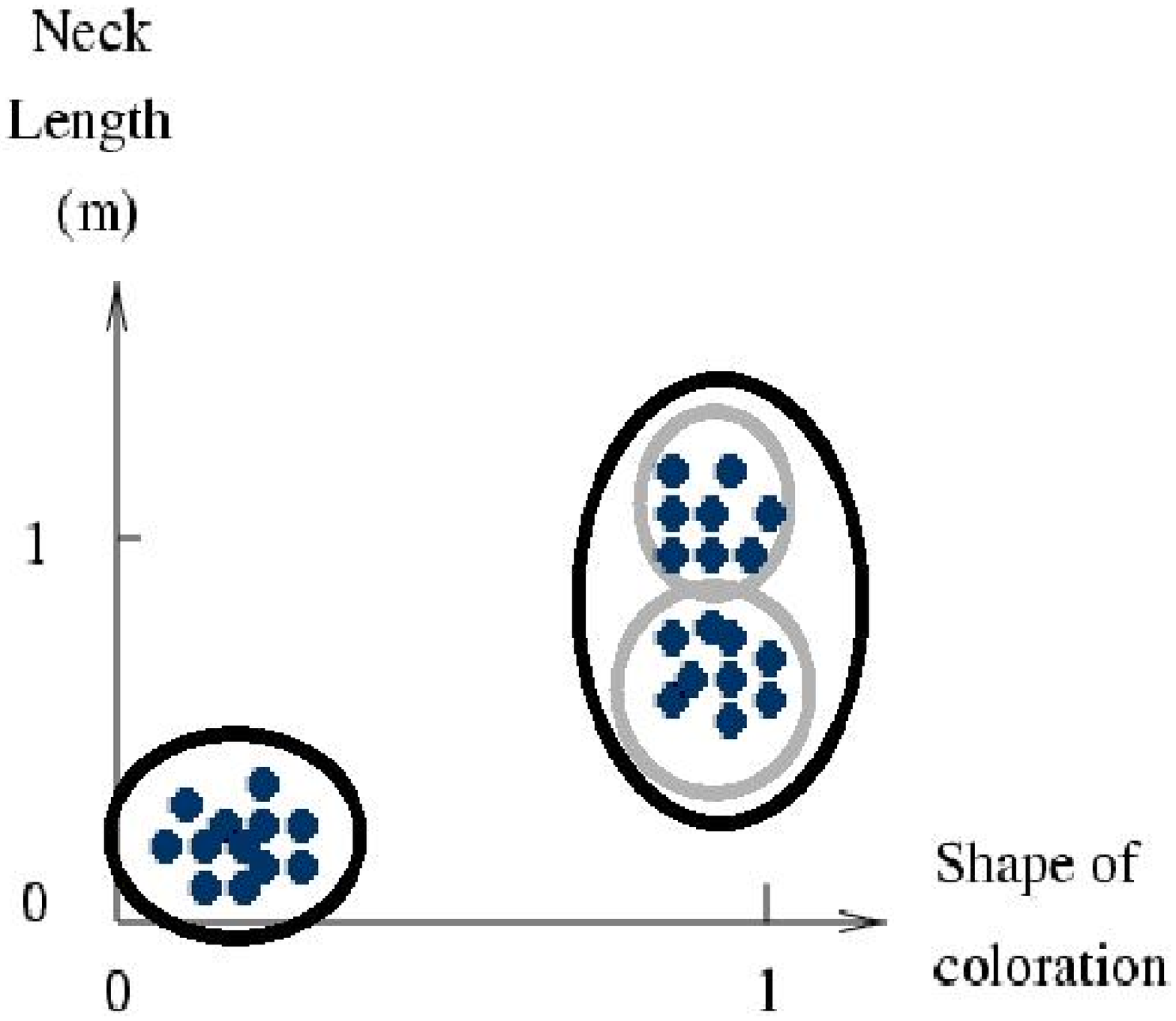,height=7cm}
    \hspace{1.0cm}
    \psfig{figure=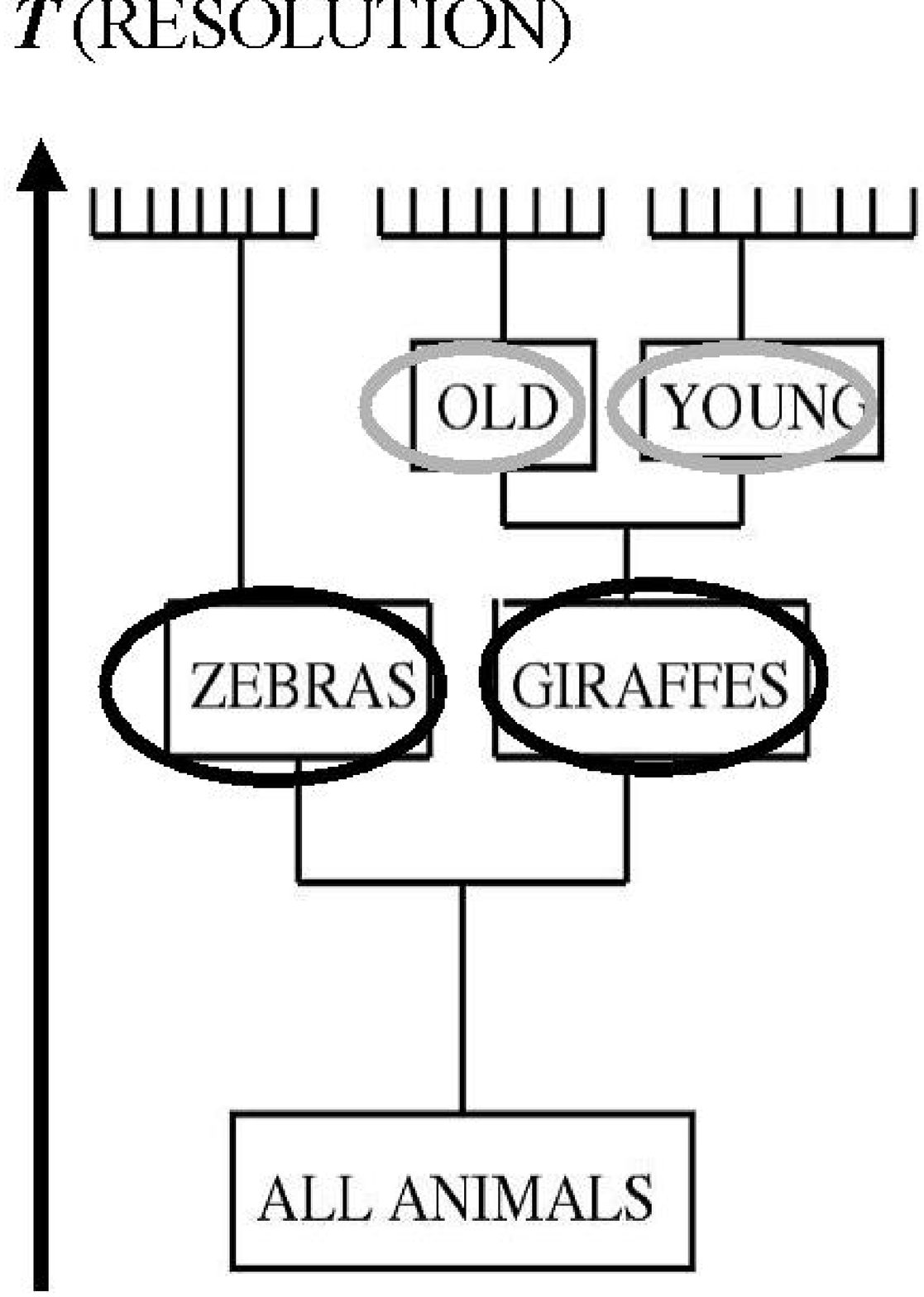,height=7cm}
    }
  \caption{Left: Each zebra or giraffe is represented as a point on the neck
  length -
  coloration shape plane. The points form two clouds marked by the black
  ellipses. At higher resolution (controlled by the parameter $T$),
  we notice that the cloud of the giraffes
  is in fact composed of two slightly separated sub clouds. The corresponding
  dendrogram is presented on the right hand side. }
  \label{fig:zebgir}
\end{figure}

Imagine the following experiment; find a child who has never seen either
a giraffe or a zebra, and expose him to a large number of pictures of these
animals without saying a word of instruction. On each animal shown the child
performs a series of $D$ measurements, two of which are most certainly  $L$,
the length of the neck, and $E$, the excentricity of the coloration
(i.e. the ratio of the small dimension and the large). Each animal
is represented, in the child's brain, as a point in a $D$ dimensional space.
Fig. \ref{fig:zebgir} depicts the projection of
these points on the two dimensional $(L,E)$ subspace. 

Even though initially the child will
see "animals" - i.e. assign all points to a single cloud - with time he will
realize (as his resolution improves) that in fact the data break into two
clear clouds; one with small values of $L$
and $E$, corresponding to the zebras, and the second - the giraffes -
with large $L$ and $E \approx 1$. The child, not having been instructed, will
not know the names of the two kinds of animals he was exposed to, but
I have no doubt that he will realize that  two
different kinds of creatures appear in the pictures. He has performed a 
clustering operation on
the visual data he has been presented with.

Let us pause and consider the data and the statements that were made. Are
there indeed two clouds in Fig \ref{fig:zebgir}? As we already said,
when the data are
seen with low resolution, they appear to belong to a single cloud of animals.
Improved resolution leads to two clouds - and closer inspection reveals that
in fact the cloud of giraffes breaks into two sub-clouds, of points that have
similar colorations but different neck lengths! Apparently there were mature
fully developed giraffes with long necks, and a group of young giraffes with
shorter necks. Finally, when resolution is improved to the level of
discerning individual differences between animals, each one forms his own
cluster. Thus the proper way of representing the structure of the data is in the
%%%%%%%%%%%%%%%%%%%%%%%%%%%%%%%%%%%%%%%%%%%%%%%%%%
form of a dendrogram, also shown in Fig \ref{fig:zebgir}.
The vertical axis corresponds to
%%%%%%%%%%%%%%%%%%%%%%%%%%%%%%%%%%%%%%%%%%%%%%%%%%%%%%
a parameter $T$ that represents the resolution at which the data are viewed.
The horizontal axis is nominal - it presents a linear ordering of the
individual data points (as identified by the
final partition, in which each cluster consists of one individual point).
The ordering is determined by the entire dendrogram - it can be thought of as a
highly nonlinear mapping of the data from $D$ to one dimension.
In any clustering algorithm that we use, we should look for the two
features mentioned here, of (a) yielding a dendrogram that starts with a single
cluster of $N$ points and ends with $N$ single-point clusters, and (b)
providing a one-dimensional ordering of the data.

\subsection{Clustering Algorithms}

There are numerous clustering algorithms. Even though each aims at achieving
a truly unsupervised and objective method, every one has built in,
implicitly or explicitly, the bias of it's inventor as to how a "cluster should
look" - e.g. a tight, spherical cloud, or a continuous region of high relative
density and arbitrary shape, etc.

Average linkage~\cite{Jain}, an agglomerative
hierarchical algorithm that joins pairs of clusters on the basis of their
proximity, is the most widely used for gene expression analysis~\cite{Eisen98}.
K-means~\cite{Jain,Duda}
and Self Organized Maps~\cite{Kohonen97} are algorithms that identify centroids
or representatives for
a preset number of groups; data points are assigned to clusters
on the basis of their distances from the centroids. There are several
physics related clustering algorithms, e.g. Deterministic
Annealing~\cite{Rose90} and
Coupled Maps\cite{Sebino}. Deterministic Annealing uses the
same cost function as K-means, but rather than minimizing it for a fixed
value of clusters $K$, it performs a statistical mechanics type analysis,
using a maximum entropy principle as its starting point. The resulting free
energy is a complex function of the number of centroids and their locations,
which are calculated by a minimization process. This minimization is
done by lowering the temperature variable slowly and following minima that
move and every now and then split (corresponding to a second order
phase transition). Since it has been proven\cite{Schneider98} that in the
generic case the free energy function exhibits first order transitions,
the deterministic annealing procedure is likely to follow one of
it's local minima.

We use another physics-motivated algorithm, which maps the clustering problem
onto the statistical physics of granular ferromagnets~\cite{Blatt96}.

\subsection{SuperParamagnetic Clustering (SPC) }

%{\bf Statistical Physics description}
The algorithm~\cite{Blatt97} assigns a Potts spin $S_i$ to each data point $i$. We use
$q=20$ components; the results depend very weakly on $q$. The distance
matrix
\begin{equation}
D_{ij}=\vert \mbox{\boldmath X} _i - \mbox{\boldmath X}_j \vert
\label{eq:Dij}
\end{equation}
is constructed. For each spin we identify a set of neighbors; a pair
of neighborings interact by a ferromagnetic\cite{Marsili01} coupling $J_{ij}=f(D_{ij})$
with a decreasing function $f$. We used a Gaussian decay, but since the
interaction between non-neighbors is set to $J=0$, the precise form of the
function has little influence on the results.

The energy of a spin configuration $\{ S \}$ is given by
\beq
{\cal H} [\{ S \}] = - \sum_{<ij>} J_{ij}[1-\delta(S_i,S_j)]
\label{eq:H}
\eeq
The summation runs over pairs of neighbors.
We perform a Monte Carlo simulation of this disordered Potts ferromagnet at
a series of temperatures. At
each temperature $T$ we measure the spin-spin correlation for every
pair {\it of neighbors},
\beq
G_{ij} = <[\delta(S_i,S_j)-1/q]/[1-1/q]>
\label{eq:Gij}
\eeq
where the brackets $< \cdot >$ represent an equilibrium average of the
ferromagnet (\ref{eq:H}), measured at $T$. If $i$ and $j$ belong to the
same ordered "grain", we will have $G_{ij} \approx 1$, whereas if the two
spins are uncorrelated, $G_{ij} \approx 0$. Hence we threshold the values
of $G_{ij}$; if $G_{ij} > 0.5$ the data points $i$ and $j$ are connected
by an edge. The clusters obtained at temperature $T$ are the connected
components of the resulting graph. In fact, the simple thresholding is
supplemented by a "directed growth" process, described elsewhere.

At $T=0$ the system is in its ground state,
all $S_i$ have the same value, and this procedure generates a single cluster
of all $N$ points. At $T=\infty$ we have $N$ independent spins,
all pairs of points are uncorrelated and the procedure yields $N$ clusters,
with a single point in each. Hence clearly $T$ controls the resolution at
which the data are viewed; as it increases, we generate a dendrogram of
clusters of decreasing sizes.

This algorithm has several attractive features, such as  (i) the number of
clusters is determined by the algorithm itself and not externally
prescribed (ii) Stability against
noise in the data; (iii) ability to identify a dense set of points,
that form a cloud of an  irregular, non-spherical shape, as a cluster.
(iii) generating a hierarchy (dendrogram) and providing
a mechanism to identify in it robust, stable clusters.

The physical basis for the last feature
is that if a cluster is made of a dense set of points on a background of
lower density, well separated from other dense regions, it will form
(become an independent magnetized grain) at a low temperature $T_1$ and
dissociate into subclusters at a high temperature $T_2$. The ratio of the
temperatures at which a cluster "dies" and "is born",
$R= T_2 / T_1$, is a measure of its stability.

SPC was used in a variety of contexts, ranging from computer
vision~\cite{Domany99} to
speech recognition \cite{Blatt97}.
Its first direct application to gene expression data
has been~\cite{Getz00} for analysis of the temporal dependence
of the expression levels in
a synchronized yeast culture~\cite{Spellman98,Eisen98},
identifying gene clusters whose variation
reflects the cell cycle.
\footnote{We have also discovered in this analysis that the
samples taken at even indexed time intervals were placed in a freezer!}
Subsequently, SPC was used~\cite{Kannan}
to identify primary targets of p53, a tumor
suppressor that acts as a trascription factor of central importance in human
cancer.

Our ability to identify stable (and statistically
significant) clusters is of central importance for our usage of SPC
in our algorithm for gene expression analysis.

\section{Clustering Gene Expression Data}
\subsection{Two way clustering}
The clustering methodology described above can be put to use for analysis
of gene expression data in a fairly straightforward way, bearing in mind the
questions and aims metioned above.

We clearly have two main seemingly distinct aims; to identify groups
of co-regulated genes which probably belong to the same machinery or
network, and to identify molecular characteristics of different
clinical states and discriminators between them.
The obvious way to go about these two tasks is by {\it Two Way Clustering}.
First view the $N$ samples as the objects to be clustered; each is represented
by a point in a $G$ dimensional "feature space", where $G$ is the number of
genes for
which expression levels were measured (in fact one works only with a subset
of the genes on a chip - those that pass some preset filters).
This analysis yields a dendrogram of samples, with each cluster containing
samples with sizeable pairwise similarities of their expression profiles
measured over the entire set of genes.

\begin{figure}[t]
%\figurebox{20pc}{15pc}{} % to have a box alone
%\epsfxsize=20pc % will enlarge or reduce the postscript figures based on the xsize
\centerline{
\psfig{figure=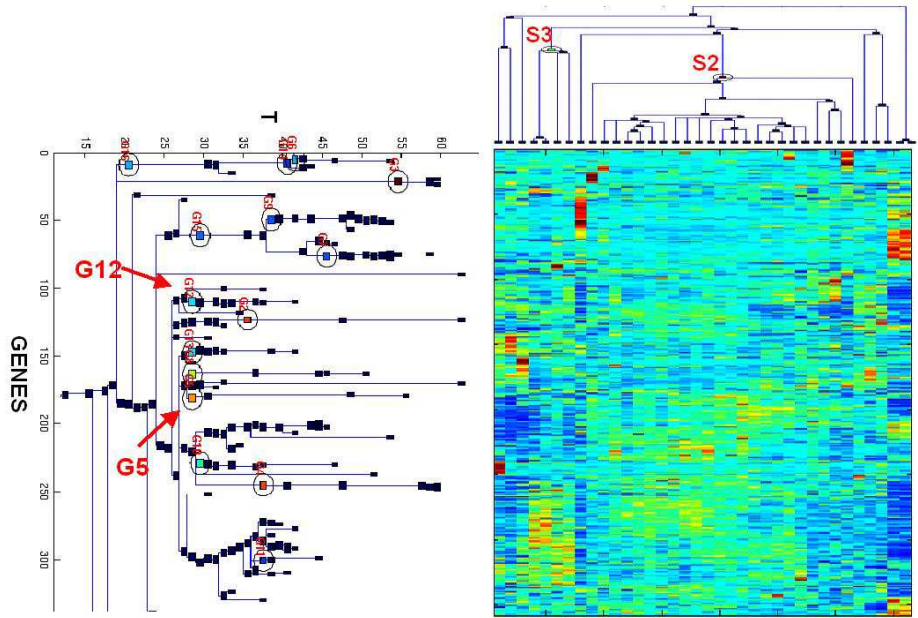,height=9cm} % postscript image file name
}
\caption{Two-way clustering of brain tumor data; the two dendrograms, of genes
and samples, are shown next to the expression matrix.  \label{fig:twoway}}
\end{figure}

The second way of looking at the same data is by considering the genes as the
objects to be clustered; $G$ data points embedded in an $N$ dimensional
feature space. This analysis groups together genes on the basis of their
correlations over the full set of samples. In Fig. \ref{fig:twoway} we present
the results of two-way clustering data obtained for 36 brain tumors (see
th enext section for details). We show here the expression matrix, with the
rows corresponding to the genes and columns to samples. The dendrograms the
correspond to the two clustering operations described above are shown next to
the matrix, whose rows and columns have been already permuted according to the
linear order imposed by the two dendrograms.

This is the type of analysis that
has been widely used in the gene expression clustering literature. It
represents a holistic approach to the problem; using every piece of
reliable information to look at the entire grand picture. This apprach does
have, however, several obvious shortcomings; overcoming these was the
motivation to develop a method which can be viewed as taking a more
reductioninst approach, while improving significantly the signal to noise
ratio of the processed data.

\subsection{Coupled Two Way Clustering - Motivation}
The main motivation of introducing CTWC~\cite{CTWC} was to
{\it increase the
signal to noise ratio} of the expression data. There are two different
kinds of "noise" the method is designed to overcome.

The first of these is a problem generated by the very advantage and most
exciting aspect of DNA-chips - the ability to view expression levels of a
very large number of genes simultaneously. Say one stays, after initial
filtering, with two thousand genes, and one wishes to study a particular
aspect of the samples (e.g. differentiating between several kinds of cancer).
Chances are that the genes which participate in the pathology of interest
constitute only a small subset of the total 2000 - say we have 40 genes
whose expression indeed distinguishes the samples on the basis of the process
that is studied. Hence the desired "signal" resides in 2 \% of the total
genes that are analysed; the remaining 98 \% behave in a way that is
uncorrelated with these and introduce nothing but noise. The contribution of
the relevant genes to
the distance between a pair of samples will be overwhelmed by the random signal
of the much larger irrelevant set. My favorite example for this situation is
that of a football stadium, in which 99,000 spectators scream at random, while
1000 others are singing a coherent tune. These 1000 are, however, scattered
all over the stadium - the chance that a listener, standing at the center
of the field, will be able to identify the tune are very small. If only we
could identify the singers, concentrate them into one stand and  point a
directional microphone at them - we could hear the signal!

In the language of gene expression analysis, we would like to
identify the relevant subset of 40 genes, and use {\it only their} expression
levels to characterize the samples. In other words, to project the datapints
representing the samples from the 2000 dimensional space in which they are
embeddded, down to a 40 dimensional subspace, and to assess the structure of
the data (e.g. - do they form two or more distinct groups?) on the basis of
this projected representation.
A similar effect may arise due to the subjects; a partition
of the genes which is much more relevant to our aims could have been obtained
had we used only a subset of the samples.

Both these examples have to do with reducing the size of the feature space.
Sometimes it is important to use the reduced set of features to cluster only
a {\it subset of the objects}. For example, when we have expression profiles
from to kinds of leukemia patients, ALL and AML, with the ALL patients breaking
further into two sub-families, of T-ALL and B-ALL, the separation of the latter
two subclouds of points may be masked by the interpolating presence of the AML
group. In other words, a special set of genes will reveal an internal structure
of the ALL cloud only when the AML cloud is removed~\cite{CTWC}.

These two statements amount to a need to work with special submatrices of the
full expression matrix. The number of such submatrices is, however, exponential
in the size of the dataset, and
the obvious question that arises is - how can one select the "right" submatrices
in an unsupervised and yet efficient way? The CTWC algorithm provides a
heuristic answer to this question.

\subsection{Coupled Two Way Clustering - Implementation}
CTWC is an iterative process, whose
starting point is the standard two way clustering mentioned above.
Denote the set of all samples by S1 and that of all genes used as G1.
The notation S1(G1) stands for the clustering operation of all samples, using
all genes, and G1(S1) for clustering the genes using all samples. From both
clustering operations we identify {\it stable clusters} of genes and samples,
i.e. those for which the stability index $R$ exceeds a critical value and whose
size is not too small. Stable gene clusters are denoted as GI with I=2,3,...
and stable sample clusters as SJ, J=2,3,... In the next iteration we use every
gene cluster GI (including I=1) as the feature set, to characterize and cluster
every sample set SJ. These operations are denoted by SJ(GI) (we clearly leave
out S1(G1)). In effect, we use every stable gen cluster as a possible "relevant
gene set"; the submatrices defined by SJ and GI are the ones we study.
Similarly, all the clustering operations of the form GI(SJ) are also carried
out. In all clustering operations we check for the emergence of partitions
into stable clusters, of genes and samples. If we obtain a new stable cluster,
we add it to our list and record its members, as well as the clustering
operation that gave rise to it. If a certain clustering operation did not give
rise to new significant partitions, we move down the list of gene and sample
clusters to the next pair.

This heuristic identification of relevant gene sets and submatrices is nothing
but an exhaustive search among the stable clusters that were generated. The
number of these, emerging from G1(S1) is a few tens, whereas S1(G1) generates
a few stable sample clusters usually. Hence the next stage
typically involves less than
a hundred clustering operations. These iterative steps stop when no new
stable clusters beyond a preset minimal size are generated, which usually
happens after the first or second level of the process.

In a typical analysis we generate between 10 and 100 interesting partitions,
which are searched for biologically or clinically interesting findings, on the
basis of the genes that gave rise to the partition and on the basis of available
clinical labels of the samples. It is important to note that these labels are
used {\it a posteriori}, after the clustering has taken place, to interpret and
evaluate the results.

\section{Applications of CTWC for gene expression data analysis}
So far CTWC has been applied primarily to analysis of data from various kinds of
cancer. In some cases we used publicly available data, with no prior contact
with the groups that did the original acquisition and analysis.
Our initial work on colon cancer~\cite{Alon99} and leukemia~\cite{Golub99}
fall in this category.

Subsequently we collaborated with a group at the University Hospital at Lausanne
(CHUV) on Glioblastoma - in this work we were involved from early in the data
acquisition stage. Our current collaborations include work on colon cancer
and breast cancer. In the latter case we worked with publicly available data,
but its choice and the challenge to improve on existing analysis came from
a biologist. We are also involved in work on leukemia and
on meiosis~\cite{Primig00} in yeast; finally, the same method
was applied successfully~\cite{Quintana02} to
analyze data obtained from an "antigen chip", used to study the antibody
repertoire of subjects that suffer from autoimmune diseases, such as diabetes.

I will limit the discussion here to presentation a few select results obtained
for glioblastoma~\cite{Godard02} and for breast cancer~\cite{ItaiMSc}.

\subsection{CTWC analysis of brain tumors (gliomas)}
Brain tumors are classified into three main groups. Low grade astrocytoma (A)
are small sized tumors at an early stage of development. Cancerous
growth may recur after their removal, giving rise to secondary gliomas (SC).
The third kind
are primary (PR) glioblastoma (GBM); this classification is assigned
when at the stage of initial diagnosis
and discovery the tumor is already of a large size.
A dataset $S1$ of 36 samples was obtained by a group from the University
Hospital at Lausanne~\cite{Godard02}. 17 of these were from PR GBM,
4 - from SC, 12 were from A and 3 from human
glioma cell lines grown in culture. Expression profiles were obtained
using Clontech Atlas 1.2 arrays of 1176 genes. For each gene $g$ the
measured expression value for tumor sample $s$ was divided by its value in a
reference sample composed of a mixture of normal brain tissue. We filtered
the genes by keeping only those for which the maximal value of this ratio
(over the 36 samples)
exceeded its minimal value by at least a factor of two. 358 genes passed this
filter and constituted our full gene set $G1$, which was clustered using
expression ratios over $S1$. The $G1(S1)$ clustering operation
(see Fig \ref{fig:twoway}) yielded 15 stable
gene clusters. The complementary operation $S1(G1)$ did not yield any partition
of the samples that could be given clear clinical interpretation.

\begin{figure}[t]
%\figurebox{20pc}{15pc}{} % to have a box alone
%\epsfxsize=15pc % will enlarge or reduce the postscript figures based on the xsize
\centerline{
\psfig{figure=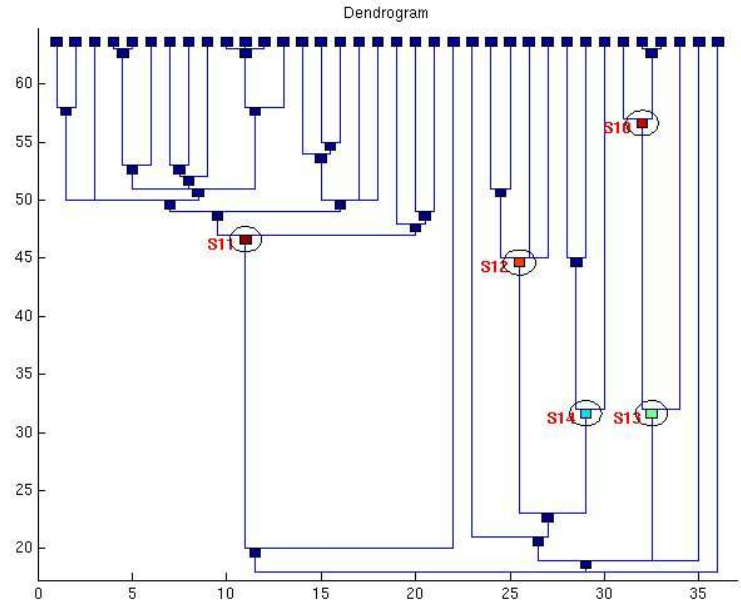,height=9cm} % postscript image file name
}
\caption{The  operation $S1(G5)$,
clustering all tumors on the basis of their expression
profiles over the genes of cluster $G5$. A stable cluster, $S11$ emerges,
containing
all the non-primary tumors and only two of the primaries.  \label{fig:s1g5}}
\end{figure}

One of the stable gene clusters, $G5$, contained
9 genes.
When the expression levels of only these genes
are used to characterize the tumors
[in the operation
denoted $S1(G5)$],  a large and stable cluster,
$S11$,  of 21 tumors, emerged (see Fig \ref{fig:s1g5}.
This cluster contained all
the 12 astrocytoma and all 4 SC tumors.
Three of the remaining 5 tumors of $S11$ were cell lines
and two were registered as
PR GBMs. Pathological diagnosis was redone for these two
tumors; one was found to contain a significant
oligoastrocytoma component, and much of the piece of the other,
that was used for RNA extraction, was diagnosed as of normal brain
ifiltrative zone. Hence the expression levels of $G5$ gave rise to
a nearly perfect separation of PR from non-PR (A and SC tumors). The
genes of $G5$ were significantly upregulated in PR and downregulated
in A and SC.

These findings made good biological sense, since
three of the genes in $G5$ (VEGF, VEGFR and PTN) are related to angiogenesis.
Angiogenesis is the process of development of blood vessels, which are
essential for growth of tumors beyond a certain critical size, bringing
nutrition to and removing waste from the growing tissue. Upregulation of 
genes that are known to be involved in angiogenesis
is a logical consequence of the fact that PR GBM are large tumors.

An important application of the method concerns
investigation of the genes that belong to $G5$; in particular, one of the
genes of $G5$, IGFBP2, was of considerable interest with little existing
clues to its function and role in cancer development. Our finding, that its
expression is strongly correlated with the angiogenesis related
genes came as a surprise that was worth detailed further study.
The co-expression of genes from the IGFBP family
with VEGF and VEGFR has been demonstrated in an
independent experiment that tested this directly for cell lines under different
conditions.

This example demonstrates the power of CTWC; a subgroup of genes with
correlated expression levels was found to be able to separate PR from
non-PR GBM, whereas using all the genes introduced noise that wiped out
this separation. In addition, by looking at the genes of this correlated
set, we provided an indication for the role that a gene with previously unknown
function may play in the evolution of tumors.

For other findings of interest in this data set we refer the reader to the
paper by Godard et al~\cite{Godard02}.

\subsection{Breast Cancer Data}
In a different study, on breast cancer, we used publicly available expression
data of Perou et al \cite{Perou00}. The choice of this particular data set
was guided by D. Botstein, who
informed us that these were of the highest quality, were submitted to most
extensive effort for analysis and challenged us to demonstrate that our method
can extract findings that eluded previous treatments. The results of this
study are available\cite{ItaiMSc}; here I present only one particular
new finding.

The Stanford data contained expression profiles of 65 human samples ($S1$)
and 19 cell lines. 40 tumors were paired, with samples taken before and after
chemotherapy (with doxorubicin), to which 3 (out of 20) subjects responded
positively. 1753 genes ($G1$) passed initial filtering; the clustering
operation $S1(G1)$, of all the samples
using their expression profiles over all these genes, did not yield any
clear meaningful partitions. Perou et al realized the same point that has
motivated us to construct CTWC, namely that one has to prune the number of
genes that are used in order to improve the signal to noise ratio. They
ranked the genes according to a figure of merit they introduced, which measures
the proximity of expressions of the two samples taken from the same patient
before and after chemotherapy, versus the (expectedly larger) dissimilarity
of samples from different patients. The 496 top scorers constituted their
"intrinsic gene set" which was then used to cluster the samples.

We did not use this intrinsic set but rather, applied CTWC on the full
sets of samples and genes. In the $G1(S1)$
operation we
found several stable gene clusters. One of these, $G46$, contained
33 genes, whose expression
levels correlate well with the cells' proliferation rates. Only 2 out of
these made it into the intrinsic set of Perou et al; hence they could not have
found any result that we obtained on the basis of these genes.

\begin{figure}[t]
%\figurebox{20pc}{15pc}{} % to have a box alone
%\epsfxsize=20pc % will enlarge or reduce the postscript figures based on the xsize
\centerline{
\psfig{figure=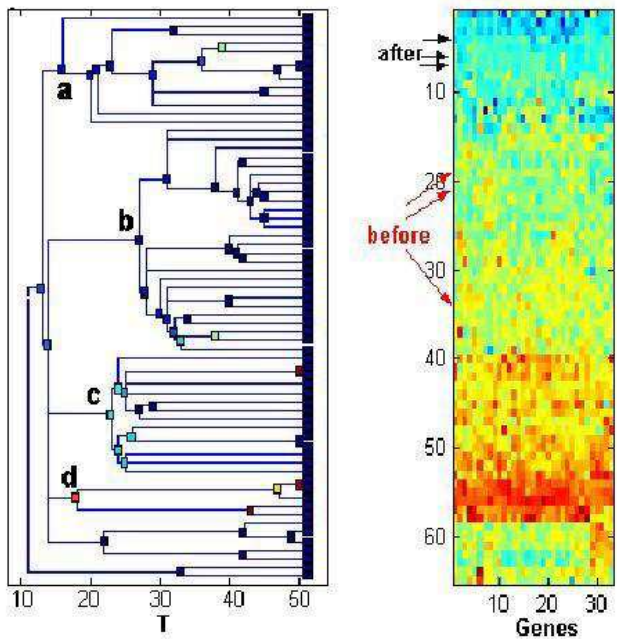,height=9cm} % postscript image file name
}
\caption{The operation $S1(G46)$, clustering all tumors on the basis
of the proliferation related genes of $G46$.  We found a cluster
(b) which contained all three samples from patients for who
chemotherapy was successful, taken {\it before} the treatment.
Cluster (b) contained 10 out of the 20 "before" samples.
\label{fig:breast}}
\end{figure}
The operation $S1(G46)$ identified three main clusters;
%Using these genes to
%cluster the 65 tumors we found the dendrogram shown in Fig. 2. The cluster
%marked 'a'
(a) of samples with low proliferation rates - these are 'normal
breast - like'; (b) samples with intermediate, and (c) with high
proliferation rates. Interestingly,
the "before treatment" samples taken from {\it all three} tumors for
which chemotherapy did succeed were in cluster (b),
whereas  the corresponding 'after treatment' samples were in (a), the
'normal breast - like' cluster. Therefore the genes of $G46$
can perhaps be used a posteriori, to indicate success of treatment on the
basis of their expression measured after treatment and, more importantly,
may have
predictive power with respect to the probability of success of
the doxorubicin therapy.
that was used. Intermediate
expression of the G46 genes may serve as a marker for a relatively high
success rate of the Doxorubicin treatment (3/10 versus 3/20 for the entire
set of "before treatment" samples). Clearly these statements are backed only by
statistics based on small samples, but they do indicate possible
clinical applications
of the method, provided experiments on more samples strengthen the statistical
reliability of these preliminary findings.

\section{Summary}
DNA chips provide a new, previously unavailable glimpse into the manner
in which the expression levels of thousands of genes vary as a function
of time, tissue type and clinical state. Coupled Two Way Clustering
provides a powerful tool to mine large scale expression data by identifying
groups of correlated (and possibly co-regulated) genes which, in turn, are
used to divide the samples into biologically and clinically relevant groups.
The basic "engine" used by CTWC is a clustering algorithm rooted in the
methodology of and insight gained from Statistical Physics.

The extracted information may enlarge our body of general basic knowledge and
understanding, especially of
gene rgulatory networks and processes. In addition, it may provide clues
about the function of genes and their role in various pathologies; one
can also hope to develop powerful diagnostic and prognostic tools based
on gene microarrays.

\section*{Acknowledgments}
I have benefited from advice and assistance of my students
G. Getz, I. Kela, E. Levine and many others. I am particularly grateful to
the community of biologists who were extremely open minded, receptive and
helpful at every stage of our entry to their fields:
D. Givol provided our first new data, as well as invaluable
advice and encouragement. The CHUV group, in particular Monika Hegi and Sophie
Godard, shared their data and knowledge generously, D. Notterman and U. Alon
were instrumental in getting us started on their colon cancer experiment,
D. Botstein
guided us towards his best breast cancer data, I. Cohen was a
powerful driving force motivating us to
apply our methods to "antigen chips" which he invented.
Our work has been supported by grants from the Germany-Israel Science Foundation
(GIF) the Israel Science Foundation (ISF) and the Leir-Ridgefield Foundation.

\end{document}